\documentclass[%
reprint,amsmath,amssymb,aps,prd]{revtex4-1}

\usepackage{graphicx}
\usepackage{dcolumn}
\usepackage{bm}
\usepackage{color}
\usepackage{float}

\newcounter{figureidx}
\stepcounter{figureidx}

\def\apj{Astrophys. J.~}

\def\apjl{Astrophys. J. Lett.~}
\def\mnras{Mon. Not. R. Astr. Soc.~}
\def\prd{Phys. Rev. D~}

\def\physrep{Phys. Rep.~}
\def\araa{Ann. Rev. Astr. \& Astrophys.~}
\def\aap{Astr.\& Astrophys.~}

\def\jcap{J. Cosmo. \& Astroparticle Phys.~}
\def\nat{Nature}

\def\msun{M_{\odot}}

\def\kms{\rm km\,s^{-1}}

\def\ent{\epsilon_{\rm nt}}

\def\ksib{\xi_{_{\rm B}}}

\def\ns{n_{\rm s}}
\def\ts{T_{\rm s}}

\def\rs{R_{\rm s}}
\def\vs{v_{\rm s}}
\def\mp{m_{\rm p}}

\def\emax{E_{\rm max}}
\def\emin{E_{\rm min}}

\def\ent{\epsilon_{\rm nt}}
\def\fkin{f_{\rm kin}}
\def\gp{\Gamma_{\rm p}}
\def\ecr{E_{\textsc{\tiny CR}}}
\def\lcr{L_{\textsc{\tiny CR}}}
\def\eg{E_{\rm g}}
\def\ep{E_{\rm p}}
\def\eth{\epsilon_{\rm th}}
\def\tsal{t_{\rm Sal}}
\def\ent{\epsilon_{\rm nt}}
\def\lkin{L_{\rm kin}}
\def\lbol{L_{\rm bol}}
\def\mh{M_{\rm halo}}
\def\Phicr{\Phi_{_{\rm CR}}}
\def\omgm{\Omega_{\rm M}}
\def\omgl{\Omega_{\Lambda}}
\def\ns{n_{\rm s}}
\def\ts{T_{\rm s}}
\def\tdyn{t_{\rm dyn}}
\def\tacc{t_{\rm acc}}
\def\tcool{t_{\rm cool}}
\def\gcmb{\gamma_{_{\rm CMB}}}
\def\tpp{t_{pp}}
\def\tpg{t_{p\gamma}}
\def\sigpp{\sigma_{pp}}
\def\kpp{\kappa_{pp}}
\def\ekin{E_{\rm kin}}

\def\eth{\epsilon_{\rm th}}

\def\sigpg{\sigma_{p\gamma}}
\def\egm{\epsilon_{\gamma}}
\def\wc{\omega_{\rm c}}
\def\esh{E_{\rm sh}}
\def\va{v_{\textsc{\tiny A}}}

\begin{document}
\preprint{APS/123-QED}
\title{Ultra high energy cosmic rays from non-relativistic quasar outflows}
\author{Xiawei Wang}
\email{xiawei.wang@cfa.harvard.edu}
\author{Abraham Loeb}
\affiliation{Department of Astronomy, Harvard University, 60 Garden Street, Cambridge, MA 02138, USA}

\date{\today}
\begin{abstract}
It has been suggested that non-relativistic outflows from quasars can naturally account for the missing component of the extragalactic $\gamma$-ray background and explain the cumulative neutrino background through pion decay in collisions between protons accelerated by the outflow shock and interstellar protons.
Here we show that the same quasar outflows are capable of accelerating protons to energies of $\sim 10^{20}$ eV during the early phase of their propagation.
The overall quasar population is expected to produce a cumulative ultra high energy cosmic ray flux of $\sim10^{-7}\,\rm GeV\,cm^{-2}s^{-1}sr^{-1}$ at $\ecr\gtrsim10^{18}$ eV. 
The spectral shape and amplitude is consistent with recent observations for outflow  parameters constrained to fit secondary $\gamma$-rays and neutrinos without any additional parameter tuning.
This indicates that quasar outflows simultaneously account for all three messengers at their observed levels.
\end{abstract}
\maketitle
%
\textit{Introduction.}---The observed ultra high energy cosmic ray (UHECR) spectrum is characterized by various spectral features \cite{hillas2006, kotera2011}.
The hardening of the spectrum at $\sim4\times10^{18}$ eV, so-called the ankle, can be produced by a transition from Galactic to extragalactic cosmic rays (CRs) for either mixed composition or iron-dominated models \cite{allard2007}, or by pair production propagation losses in proton-dominated models \cite{berezinsky2006}.
The flux suppression detected above $\sim3\times10^{19}$ eV, is either caused by the interaction between UHECRs and the cosmic microwave background (CMB) photons, the so-called Greisen-Zatsepin-Kuzmin (GZK) cutoff \cite{greisen1966, zatsepin1966}, or is potentially associated with the maximum energy of the accelerated nuclei \cite{aloisio2011}.
The spectrum can be fitted by a power-law with spectral index of $\sim 3$ between the cosmic knee ($\sim 10^{15}$ eV) and the ankle, and $\sim 2.6$ between the ankle and the GZK cutoff.
The origin of UHECRs remains uncertain but it is believed to be of an extragalactic origin \cite{kotera2011}.

Growing observational evidence reveals the existence of large-scale outflows driven by the active galactic nuclei (AGN).
It includes the detection of multi-phase outflows in nearby ultraluminous infrared galaxies \cite{rupke2011, tombesi2015} and the presence of broad absorption lines in quasars  \cite{zakamska2014, arav2015}.
In previous work \cite{wang2015}, we derived a detailed hydrodynamical model of quasar outflow's interaction with the ambient interstellar medium (ISM) (See Ref.\cite{wang2015} for details).
Protons accelerated by the outflow shock to relativistic energies interact with the interstellar protons and produce secondary $\gamma$-ray photons and neutrinos via pion production that naturally account for the missing component of the extragalactic $\gamma$-ray background (EGB) \cite{wang2016a}, as well as the cumulative neutrino background (CNB) \cite{wang2016b}.

In this Letter, we calculate the cumulative UHECR flux above $\sim 10^{18}$ eV produced by non-relativistic quasar outflows and discuss the multi-messenger implications with secondary $\gamma$-rays and neutrinos simultaneously generated by the same population of sources.
%
%

\textit{UHECR production.}---Ultra-fast winds with a velocity $\sim 0.1\, c$ are continuously injected into the ISM of the host galaxy during the quasar's lifetime \cite{king2015}, taken to be the Salpeter time $\tsal\sim 4\times10^7$ yrs, and drive a forward outflow shock that accelerates protons to relativistic energies via the Fermi acceleration, in analogy with supernova-driven shocks \cite{caprioli2012}.
Here we consider the non-relativistic spherical outflows, rather than the collimated relativistic jets seen in only $\sim 10\%$ of the AGN population \cite{fgq2012, king2015}.
The resulting proton spectrum can be described by a power-law profile with an exponential cutoff \cite{caprioli2012}:
\begin{equation}
\frac{dN}{d\ep}=N_0\ep^{-\gp}\exp\left(-\frac{\ep}{\emax}\right)\;,
\end{equation}
where $\ep$ is the proton energy, $\emax$ is the maximum energy of the accelerated protons and $\gp$ is the power-law index.
$N_0$ is the normalization constant that can be constraint by:
\begin{equation}
\int_{\emin}^{\emax} E\frac{d\dot{N}}{dE} dE=\ent\lkin\;,
\end{equation}
where the minimum proton energy $\emin\sim \mp c^2$, $\mp$ is the proton mass and $\ent$ is the fraction of outflow's kinetic luminosity $\lkin$ converted to accelerated protons.
We assume that $\lkin$ is a fraction, $\fkin$, of the quasar's bolometric luminosity $\lbol$.
Secondary $\gamma$-ray photons and neutrinos are produced via pion decay from interaction between accelerated protons and ambient protons in the ISM.
We adopt $\ent\sim0.1$ similarly to the conditions in supernova remnants (SNRs) \cite{caprioli2012, caprioli2014a} and $\fkin\sim1-5\%$ from fitting the resulting $\gamma$-rays and neutrinos to the EGB \cite{wang2016a} and CNB \cite{wang2016b}, consistently with observations \cite{ackermann2015} and theoretical models \cite{caprioli2012} of supernova shocks.
The maximum energy of the accelerated protons, $\emax$, can be extrapolated from $\emax\approx\esh\wc \tdyn/3\kappa$ for shocks with an Alfv\'en Mach number $\mathcal{M}\gtrsim100$ \cite{caprioli2014a}, where $\esh=\mp\vs^2/2$, $\wc=eB_0/\mp c$, $\kappa\propto B_0/B\propto1/\sqrt{\mathcal{M}}$, and $B_0$ and $B$ are the pre-shock and post-shock magnetic field, respectively.
$B_0$ can be obtained from equipartition of energy in the ambient ISM.
Here $\mathcal{M}=\vs/\va$, $\va=B_0/\sqrt{4\pi n_0\mp}$ and $n_0$ and $T_0$ are the ambient ISM number density and temperature, described by Ref. \cite{wang2015}.
The dynamical time, $\tdyn\sim\rs/\vs$, where $\rs$ and $\vs$ are the radius and velocity of the outflow, respectively, as determined from outflow hydrodynamics (see Ref. \cite{wang2015} for details).
For $\rs\lesssim200$ pc, $\mathcal{M}\sim 10^2-10^3$.
We can also derive $\emax$ by equating the acceleration timescale \cite{blandford1987}, $\tacc\sim\ep c/eB\vs^2$, to the minimum of the dynamical ($\tdyn$) and cooling ($\tcool$) timescale.
For simplicity, we adopt the most optimistic assumption of energy equipartition\cite{bustard2016}, in analogy to SNRs, namely that a fraction of the post-shock thermal energy is carried by the magnetic field, $B^2/8\pi=\ksib\ns k \ts$, where $\ksib\sim0.1$ based on observations \cite{chevalier1998}, $k$ is the Boltzmann constant, and $\ns$ and $\ts$ are the number density and temperature of the shocked medium, respectively.
We have verified that the results from the above two approaches are consistent.

Accelerated protons lose energies via hadronuclear ($pp$) or photohadronic ($p\gamma$) interactions.
In the $pp$ scenario, the cooling timescale is given by \cite{kelner2006}:
\begin{equation}
\tpp^{-1}=\ns\sigpp c \kpp\;,
\end{equation}
where $\kpp\sim0.5$ is the inelasticity parameter, $\sigpp\approx30[0.95+0.06\ln(\ekin/\rm 1 GeV)]$ mb is the cross section of $pp$ collision \cite{aharonian2000} and $\ekin=\ep-\mp c^2$.
The $p\gamma$ cooling timescale can be obtained by \cite{stecker1968, waxman1997}:
\begin{equation}
\tpg^{-1}=\frac{c}{2\gamma_{\rm p}^2}\int_{\eth}^{\infty}d\epsilon\,\sigpg(\epsilon)\kappa(\epsilon)\epsilon\,\int_{\epsilon/2\gamma_{\rm p}}^{\infty}d\egm\,\egm^{-2}\,n(\egm)\;,
\end{equation}
where $\eth\sim145$ MeV is the threshold energy for pion production in the rest frame of the protons and $\gamma_{\rm p}=\ep/m_{\rm p} c^2$.
The numerical approximation for the total photohadronic cross section, $\sigpg$, is taken from M{\"u}cke et al. (2000) \cite{muecke2000}.
$n(\egm)$ is the number density of soft photons in the energy range $\egm$ to $\egm+d\egm$.
We adopt a template for quasar's spectral energy distribution which includes infrared emission from the dusty torus, optical and UV emission from the accretion disk and X-ray emission from the corona \cite{marconi2004, collinson2016}.
A comparison of the relevant timescales is shown in Figure 1.
\begin{figure*}
\includegraphics[angle=0,width=1.5\columnwidth]{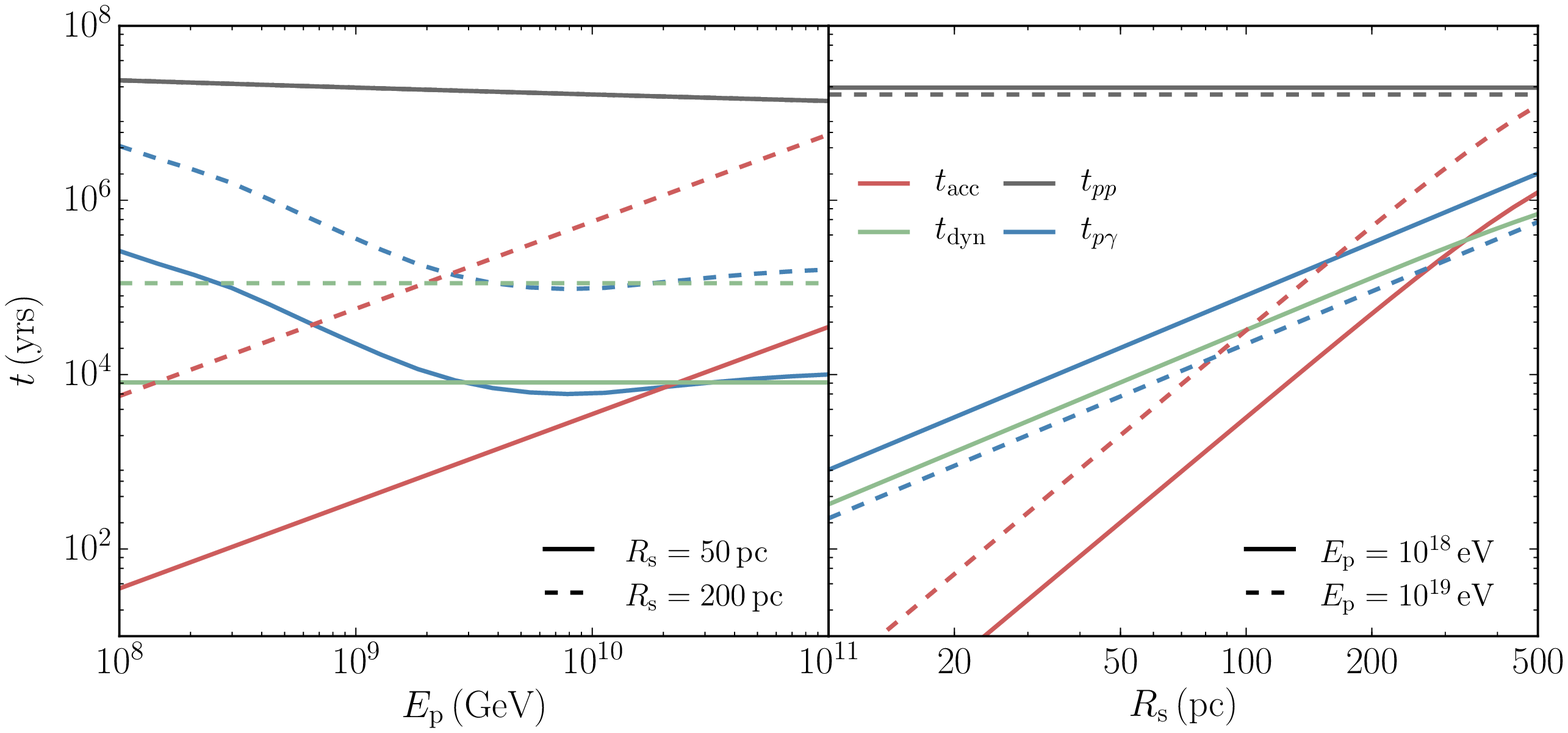}
\caption{
Comparison of relevant timescales.
On the left panel, we compare the acceleration, dynamical, $pp$ and $p\gamma$ timescales as a function of proton energy when the outflow propagates to 50 pc (solid) and 200 pc (dashed), respectively, within a host galaxy halo of mass of $10^{12}\msun$ at a redshift of $z=0.1$.
In the right panel, we show the timescales as a function of outflow radius for $\ep=10^{18}$ eV (solid) and $10^{19}$ eV (dashed).
The gas density profile is self-consistently determined by the halo mass and redshift \cite{wang2015}.
The magnetic field energy density is estimated to be a fraction $\ksib\sim0.1$ of the equipartition value.
For $\ent\sim10\%$ and $\fkin\sim5\%$,
we find that $pp$ collision timescale, $\tpp$, is substantially longer than $p\gamma$ interaction timescale, $\tpg$, at lower energies and smaller outflow radii.
Therefore, the dynamical timescale $\tdyn$ and $\tpg$ set a tighter constraint on $\emax$.
}
\end{figure*}

The most effective acceleration of UHECRs occurs in the early phase of outflow's propagation.
We estimate the optical depth of protons interacting with soft photons from the quasar and verify that only absorption of CMB photons have a non-negligible impact on the UHECR spectrum.
The resulting $\emax$ and $B$ as a function of outflow radius $\rs$ and elapsed time $t$ is depicted in Figure 2.

\begin{figure}
\includegraphics[angle=0,width=0.95\columnwidth]{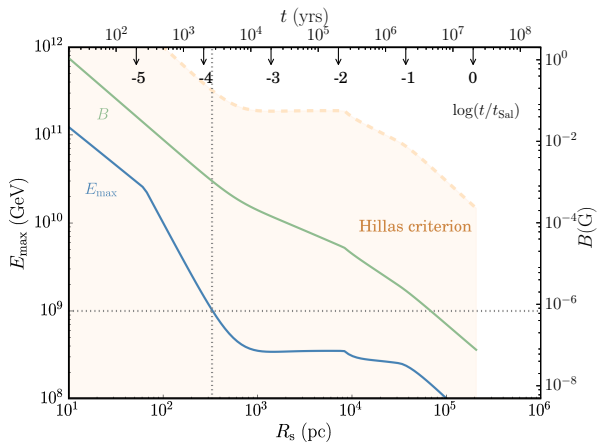}
\caption{Maximum energy of the accelerated protons $\emax$ (blue line; left vertical axis) and magnetic field behind the outflow shock $B$ (green line; right vertical axis) as a function of outflow radius $\rs$ (bottom axis) and time elapsed $t$ (top axis). 
Here, we calibrate $\emax$ and $B$ by the consideration of equipartition with the post-shock thermal energy,
for a halo mass $\mh=10^{12}\msun$, redshift $z=0.1$, $\ksib\sim0.1$, $\ent\sim10\%$ and $\fkin\sim5\%$.
The gray dashed lines mark the energy threshold of UHECRs at $E\sim10^{18}$ eV.
The upper axis is also scaled to the Salpeter time $\tsal$, indicating the fraction of a quasar's lifetime spent at each location.
The shaded beige region represents the allowed $\emax$ constrained by the Hillas criterion to confine protons \cite{hillas2006, kotera2011}.
}
\end{figure}

Figure 2 shows that $\emax$ reaches $\sim 10^{20}$ eV after the wind is launched and rapidly declines to $\lesssim10^{17}$ eV as $\vs$ decreases when the outflow enters the galactic halo, below the energy range of interest here.
The duration of UHECR production is $\sim10^{4}$ yrs, $\sim 0.01\%$ of a quasar's lifetime.
This suggests that only $\sim 0.01\%$ of quasars at any given time produce UHECRs; this sets a threshold on the sample size of AGNs needed to obtain a meaningful cross-correlation signal with the arrival directions of UHECRs.
An additional constraint on UHECR production is the size of the source and the magnetic field intensity calibrated by equipartition with the post-shock thermal energy, known as the Hillas criterion \cite{hillas1984}.
The UHECR source should be capable of confining the particles up to $\emax$, or equivalently, the size of the source must be larger than the maximum Larmor radius of the particle.
Measurements by the Pierre Auger Collaboration favor a heavier composition at the highest energies \cite{aab2015}. 
However, there are uncertainties in the modelling of hadronic interactions in the shower \cite{kotera2011}.
For simplicity, we adopt a proton-only prescription for the UHECRs accelerated by outflows since the ISM is mainly composed of protons, but we expected heavier nuclei to be accelerated as well based on the ISM metallicity.
We verified that the size of the outflow satisfies $\rs\gtrsim\ep/eB$, as shown in the shaded region of Figure 2, and find that $\tdyn$ and $\tpg$ set a tighter constraint on $\emax$.
%
%

\textit{Cumulative UHECR intensity.}---The UHECRs interact with CMB photons in the intergalactic medium and produce secondary particles via photohadronic interaction which leads to pion production, $p+\gcmb\rightarrow n+\rm pions$, and pair production, $p+\gcmb\rightarrow p+e^++e^-$.
We follow the detailed prescription given by Berezinsky et al. (2006) \cite{berezinsky2006} to calculate the corresponding energy losses, which produces the dip at $10^{18}-10^{20}$ eV, where the second flattening at $\sim 10^{19}$ eV accounts for the ankle \cite{berezinsky2006, kotera2011}.
The expected spectral shape is identical to the injection spectrum at each snapshot during the propagation of the outflow as UHECRs with energies $\gtrsim10^{18}$ eV are not confined in the Galaxy and thus propagation effect can be neglected \cite{kotera2011}.
The piling up of spectra at each outflow snapshot makes the cumulative spectrum steeper due to the decrease of $\emax$ at large $\rs$.
We estimate the cumulative UHECR intensity by summing over the entire quasar population:
\begin{equation}
\begin{split}
\ecr^2\Phicr=\frac{c}{4\pi H_0} \iint & \phi(\lbol,z)\frac{\lcr(\ecr^{\prime},\lbol,z)}{E(z)}\,\\
&\times f(\ecr^{\prime},z)\,d\log\lbol\,dz\;,
\end{split}
\end{equation}
where $\lcr=\ecr^2d\dot{N}/d\ecr$, $\ecr^{\prime}=(1+z)\ecr$ is the intrinsic CR energy, $\lbol$ is the bolometric luminosity, $\phi(\lbol,z)$ is the bolometric luminosity function of quasars \cite{hopkins2007} and $E(z)=\sqrt{\omgm(1+z)^3+\omgl}$. 
We adopt the standard cosmological parameters, $H_0=70\,\kms\rm Mpc^{-1}$, $\omgm=0.3$ and $\omgl=0.7$ \cite{ade2016}.
$f(\ecr^{\prime},z)$ is the modification factor due to interaction with the CMB photons \cite{berezinsky2006}.
We assign outflows to all quasars, consistently with the source redshift evolution rate limits set by the \textit{Fermi}-LAT and IceCube observations \cite{murase2016b}.

\begin{figure*}
\includegraphics[angle=0,width=1.9\columnwidth]{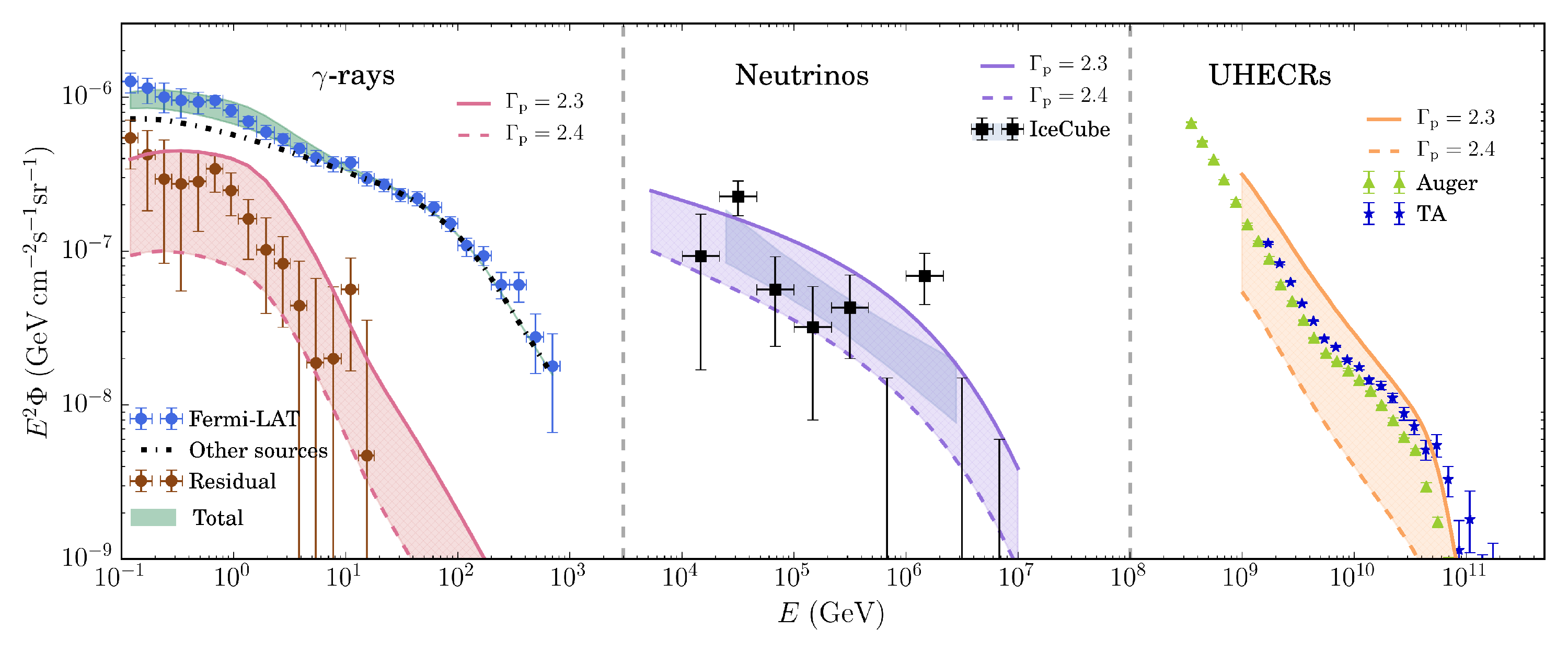}
\caption{$\gamma$-ray photons, neutrinos and UHECRs produced by quasar outflows.
From left to right, we show the cumulative $\gamma$-ray, neutrino background and UHECR flux for $\gp=2.3$ (solid line) and $\gp=2.4$ (dashed line), represented by the hatched regions, respectively.
For the $\gamma$-ray background, the contribution from other components to the EGB including blazars, radio galaxies and star-forming galaxies is plotted in comparison with the most recent \textit{Fermi}-LAT data \cite{ackermann2015}.
The cumulative neutrino background observed by IceCube \cite{aartsen2015}, represented by the data points and the gray band.
In the right section, we show the most recent data from Pierre Auger Observatory \cite{aab2015} and TA \cite{fukushima2015}, and derive the cumulative UHECR intensity without additional parameter tuning.
For simplicity, we assume a pure-proton prescription consistent with the composition of the ISM.
We find that quasar outflows naturally explain the spectra of all three messengers with parameters consistent with observations \cite{tombesi2015} and theoretical models for supernova-driven shocks \cite{caprioli2012}.
}
\end{figure*}

In Figure 3, we show the most recent $\gamma$-ray data from \textit{Fermi}-LAT \cite{ackermann2015}, neutrino data from IceCube \cite{aartsen2015} and UHECR data from the Pierre Auger Observatory \cite{aab2015} and Telescope Array (TA) \cite{fukushima2015}.
Using values of $\ent$, $\fkin$ and $\gp$ constrained by fitting $\gamma$-rays to the EGB  \cite{wang2016a} (left section) and neutrinos to the CNB \cite{wang2016b} (middle section), we derive the UHECR spectrum (right section) with $\gp\sim 2.3-2.4$ at $\ecr\gtrsim10^{18}$ eV without additional parameter tuning.
For $\ent\sim10\%$, the best fit $\fkin\sim1-5\%$ \cite{wang2016a} is consistent with theoretical models and observations \cite{caprioli2012, tombesi2015}.
It is important to note that we naturally obtain the spectral shape and amplitude of the UHECR flux from the same outflow  model that explains the EGB and CNB.
A simultaneous fit to the UHECR spectrum, composition and anisotropy is challenging, as shown by the preliminary results from the Pierre Auger Collaboration \cite{aab2015}.
The spectrum could be sensitive to the detailed photohadronic interactions during UHECR propagation \cite{batista2015}, while the spectral shape might be affected by the presence of intervening magnetic fields at $\ecr\lesssim10^{18}$ eV \cite{batista2014}.
%
%

\textit{Multi-messenger implications.}---Secondary photons and neutrinos are produced as UHECRs interact with the ambient interstellar protons.
The resulting $\gamma$-ray photons can naturally account for the missing component of the EGB at $\eg\lesssim10$ GeV as suggested by the most recent \textit{Fermi}-LAT observation \cite{ackermann2015, wang2016a}, while the associated neutrinos explain the CNB as observed by IceCube \cite{aartsen2015, wang2016b}.
With $\ent\sim10\%$, $\fkin\sim1-5\%$ and $\gp\sim2.3-2.4$, constrained to fit the \textit{Fermi}-LAT and IceCube data, we naturally explain the UHECR flux without additional parameter tuning, as shown in Figure 3.
This is consistent with parameter values inferred from observations of outflows \cite{tombesi2015} as well as the branching ratio between secondary $\gamma$-rays and neutrinos, which sets an upper limit on the power-law index of the injection spectrum to be $\lesssim 2.2-2.4$ \cite{wang2016b, murase2016}.
Indeed, recent $\gamma$-ray observations suggest the existence of hadronic emission from an outflow in a nearby galaxy \cite{lamastra2016}.
However, the predicted $\gamma$-ray emission from an individual outflow is too faint to be detected outside the local Universe ($z\sim0.1$), explaining why these outflows have been barely detected in $\gamma$-rays.
The simultaneous radio emission from accelerated electrons by the same outflow shocks is sufficiently bright to be observed to a redshift of $\sim5$ and is free of contamination from scattered quasar light by the surrounding electrons in the halo \cite{wang2015}. 
Radio observations with the \textit{Jansky Very Large Array} and the \textit{Square Kilometre Array} could therefore directly image the shock front.
Stacking analysis of $\gamma$-rays and neutrinos can be performed in the future to search for more direct evidence of quasar outflows \cite{wang2016a}.
Alternative UHECR sources such as blazars \cite{murase2012} could make up to $\sim50\%$ of the EGB at $\eg\lesssim10$ GeV through synchrotron self-Compton emission and potentially dominate the EGB at higher energies \cite{ajello2015}.
However, they produce only $\sim10\%$ of the CNB at energies below $\sim0.5$ PeV  \cite{padovani2015}.
Radio galaxies with misaligned jets can accelerate UHECRs via the same mechanism as blazars \cite{dermer2009}. 
However, they account for only $\lesssim10\%$ of the EGB at $\eg\lesssim10$ GeV \cite{wang2016a} and do not fully account for the CNB.
Another potential UHECR source is the gamma-ray bursts (GRBs) \cite{waxman1997}, which can not account for most of the EGB.
The identification of UHECR sources with $\gamma$-ray and neutrino sources would provide a smoking gun evidence for their origin \cite{becker2005, kotera2011}.
%
%

\textit{Summary.}---In this Letter, we have shown that the cumulative UHECR flux produced by non-relativistic quasar outflows naturally accounts for the observed spectrum at $\ecr\gtrsim10^{18}$ eV by Auger \cite{aab2015} and TA \cite{fukushima2015}.
We constrained the free parameters of the model to fit data on the secondary $\gamma$-rays and neutrinos without additional parameter tuning.
We find that the best fit power-law index of the injection spectrum is $\gp\sim2.3-2.4$, consistent with observations of supernova remnants and theoretical models \cite{caprioli2012}.
Altogether, quasar outflows simultaneously produce all three messengers -- $\gamma$-rays, neutrinos and UHECRs -- that account for the missing component of the EGB, the CNB and the observed UHECR spectrum.
%

We thank Xuening Bai and Rafael Alves Batista for helpful comments on the manuscript.
This work is supported by NSF grant AST-1312034.

%
\end{document}